\documentclass[aps,epsf,subfigure,twocolumn]{revtex4}

\usepackage{graphicx}
\usepackage{epstopdf}
\usepackage{dcolumn}
\usepackage{bm}
\usepackage{amsmath}
\usepackage{subfigure}
\usepackage{color,psfrag,epsfig}
\usepackage{multirow}
\usepackage{float}

\begin{document}

\newcommand{\up}{{\mid \uparrow \rangle}}
\newcommand{\down}{{\mid \downarrow \rangle}}

\title{Non-homogeneity of the density of states of tunneling two-level systems at low energies}

\author{A. Churkin$^{1,2}$}
\author{D. Barash$^2$}
\author{M. Schechter$^{1,^*}$}
\affiliation{$^1$Department of Physics, Ben Gurion University of
the Negev, Beer Sheva 84105, Israel}
\affiliation{$^2$ Department of Computer Science, Ben Gurion University of
the Negev, Beer Sheva 84105, Israel}


\date{today}

\begin{abstract}

Amorphous solids, and many disordered lattices, exhibit a remarkable qualitative and quantitative universality in their acoustic properties at temperature $\lesssim 3$K. This phenomenon is attributed to the existence of tunneling two-level systems (TTLSs), characterized by a homogenous density of states (DOS) at energies much lower than the disorder energy ($\approx 0.1$eV).
Here we calculate numerically, from first principles, the DOS of KBr:CN glass, the archetypal disordered lattice showing universality. In contrast to the standard tunneling model, we find that the DOS diminishes abruptly at $\approx 3$K, and that tunneling states differ essentially by their symmetry under inversion. This structure of the TTLSs dictates the low temperature below which universality is observed, and the quantitative universality of the acoustic properties in glasses. Consequences to the properties of glasses at intermediate temperatures, as well as to the microscopic structure of amorphous solids, are discussed.

\end{abstract}

\pacs{61.43.Fs, 66.35.+a, 61.72.S-, 63.20.kp}

\maketitle

\section {Introduction}
\label{Sec:Introduction}

The existence of two-level tunneling defects as a generic property in amorphous systems was postulated four decades ago\cite{AHV72,Phi72} in an attempt to explain the remarkable universality in the low energy characteristics of amorphous solids as were found earlier by Zeller and Pohl\cite{ZP71}. The exact nature of the tunneling states is not known. Yet, their characteristics at low energies has been thoroughly studied through measurements of properties such as specific heat, thermal conductivity, and internal friction (see Refs. \cite{HR86,Phi87} and references therein), as well as their
relaxation and dephasing times
via echo experiments (see e.g. \cite{GG76,GSHD79,NFHE04}). At very low temperatures interactions between the tunneling two-level systems (TTLSs) lead to glassy characteristics and slow relaxation\cite{Bur95,RNO96}. Recently, in remarkable experiments the coupling between superconducting qubits and TTLSs was used to study single TTLSs, their relaxation times, and the dependence of their bias energy on applied strain\cite{LSC+10,GPL+12}.

Many of the above mentioned properties can be explained using what is now referred to as the "standard tunneling model"(STM)\cite{AHV72,Phi72,Jac72}.
In the STM each TTLS is characterized by the energy bias between its two states $\Delta$, the tunneling amplitude between the two states $\Delta_o$, and its coupling to the phonon field denoted by $\gamma$. A central assumption of the STM for the ensemble of TTLSs in a given system is that the energy biases and the magnitudes of the barriers are homogenously distributed, resulting in the distribution $P(\Delta,\Delta_o) = P_o/\Delta_o$. The STM further assumes that $\gamma$ has a narrow distribution.
With regard to the energy biases, the assumption of homogeneity at low energies rests on the argument that the biases are dictated by the large energy related to the disorder, $\approx 0.1$eV, and are therefore homogenous at much lower energies.

Although very successful in explaining many of the universal properties at low temperature, the STM falls short in describing the nature of the tunneling states, the quantitative universality of phonon attenuation, and the energy scale of $\approx 3$K, only below which universality is observed. As the STM has no energy scale except that of the glass transition, it cannot also explain the drastic change of behavior above $\approx 3$K, and phenomena such as the universal plateau in thermal conductivity at $3-10$K, and the boson peak at higher energies\cite{BND84,KB11}. Thus, the mechanism, generic to the disordered state, that leads to the universality of phonon attenuation at low temperatures, and its relation to the microscopic structure of the amorphous state, remain long standing open questions.

The low energy phenomena dictated by the TTLSs are observed equally
in disordered lattices\cite{PLT02}. It was carefully shown\cite{YKMP86,LVP+98} that the phenomena are equivalent between the two systems,
strongly suggesting that it is the same mechanism leading to the universal phenomena in all disordered systems. It is therefore expected that the central characteristics of the low energy excitations, such as the distribution of their coupling constants and the structure of their density of states (DOS), will be similar between the different systems showing universality. It was thus concluded\cite{LVP+98,PLC99} that the defects in the crystal should be used to model the excitations in amorphous solids, rather than the amorphous structure itself.

In this paper we present numerical results for the density of tunneling states in KBr$_{1-x}$(CN)$_x$, where the system is modeled by its bare inter-atomic potentials. KBr$_{1-x}$(CN)$_x$ is the archetypal disordered lattice showing universality at concentrations $0.2<x<0.7$, and the magnitude of its energy disorder at the relevant concentration range is $\approx 0.05$eV, similar to that of amorphous solids. Thus, we could expect a homogenous TTLS DOS for energies smaller than $\approx 500$K also in KBr:CN.

Using a combined Monte Carlo (MC) and molecular statics technique, we relax fragments of KBr:CN in both 2 dimensions (2D) and 3 dimensions (3D) to a low energy state. We then calculate the energy of all states resulting from a tunneling of a single CN defect. Remarkably, a sharp decrease of the DOS of the tunneling states is found at energy $\approx 3$K in both 2D and 3D. This energy is much smaller than the energy scale of the disorder, as is reflected e.g. in the glass temperature. Yet, it is the same energy scale that dictates the temperature below which universality is observed. We further find that it is CN flips, which constitute two states related to each other by local inversion symmetry, that overwhelmingly dominate the low energy excitations. At the same time, CN rotations (asymmetric with respect to local inversion) have a much larger energy scale ($\approx 500$K in 2D, $\approx 300$K in 3D), and a diminishing (non-homogenous) DOS at low energies.

Our results here add up to the bimodality of the coupling strengths of TTLSs to the phonon field, as was obtained numerically in \cite{GS11}. Both these findings are in sharp contradiction to the assumptions of the STM, but in agreement with a recent ''two-TTLS" theory\cite{SS09} for the low temperature universality.
Our results thus suggest that it is the inhomogeneity of the DOS of TTLSs at low energies, and its dependence on the symmetry of the TTLSs, that
dictates quantitative universality of the acoustic properties of glasses, and the energy scale related to it.

\section{Numerical procedure}
\label{Sec:Numerical}

For a given realization of CN impurities, the low energy state of a KBr$_{1-x}$(CN)$_x$ sample corresponds to a specific orientation of each of the CN impurities in the sample, which is dictated by the impurity-impurity interactions. At the same time, the exact positions of all ions in the sample giving energy minimization have to be calculated. We therefore use a technique that hybridizes local energy minimization (conjugate gradients) and MC simulation.

We start by creating a 3D grid of volume $N \times N \times N$  ($N \times N$ in 2D, $N$ is even)  of $K$\textsuperscript{+} and $Br$\textsuperscript{-} ions, having distance $3.1974\AA$  ($3.2735\AA$ in 2D)  between the ions. These distance values were calculated by the energy minimization procedure of pure KBr grid. We replace randomly some of the $Br$\textsuperscript{-} ions by $CN$\textsuperscript{-} ions, according to our chosen concentration $x=0.25$. The charges of $K$\textsuperscript{+} and $Br$\textsuperscript{-} ions are taken as +1 and -1 respectively, and the charge of the $CN$\textsuperscript{-} ion is represented by fractional charges  q$_{C}=-1.28$, q$_{N}=-1.37$ and q$_{center}=+1.65$ placed respectively on the carbon atom, nitrogen atom and at the center of mass \cite{KM83}. The distance of the carbon and nitrogen atoms from the center of mass are taken as $0.63\AA$ and $0.54\AA$ respectively \cite{KM83}.
Interatomic potential is calculated by the formula:

\begin{equation}
V_{\alpha \beta}(R) = A_{\alpha \beta}\exp(-a_{\alpha \beta}R) + {B_{\alpha \beta} \over R^6} + K{q_{\alpha}q_{\beta} \over R}.
\label{potential}
\end{equation}

The interatomic potential parameters A$_{\alpha \beta}$, $a_{\alpha \beta}$ and B$_{\alpha \beta}$, see Table~\ref{tab:params}, are taken from Ref. \cite{BK82},
and $K=1389.35 \AA$kJ/mol.

\begin{table}
\def\arraystretch{1.5}
\begin{tabular}{|c|c|c|c|}
	\hline
$_{\alpha \alpha}$ &  $A_{\alpha \alpha}$ (kJ/mol)  &  $a_{\alpha \alpha}$ (1/$\AA$) & $B_{\alpha \alpha}$ ($\AA^6$kJ/mol)   \\
	\hline
KK & 158100 & 2.985 & -1464 \\
CC & 259000 & 3.600 & -2110 \\
NN & 205020 & 3.600 & -1803 \\
BrBr & 429600 & 2.985 & -12410 \\
	\hline
\end{tabular}
\caption {Interatomic potential parameters (taken from Ref. \cite{BK82}). Cross-interaction parameters were calculated by $A_{\alpha \beta}=(A_{\alpha \alpha}A_{\beta \beta})^{1/2}$, $a_{\alpha \beta}=(a_{\alpha \alpha}a_{\beta \beta})/2$, $B_{\alpha \beta}=-(B_{\alpha \alpha}B_{\beta \beta})^{1/2}$. }
\label{tab:params}
\end{table}

For a single CN impurity in an otherwise pure KBr lattice we find that its preferred orientation is along the eight-fold degenerate in-space diagonals, in agreement with Ref.~[\onlinecite{Bey75}]. In two-dimensions the degeneracy is four-fold, along the plane diagonals. These degeneracies are lifted by tunneling. However, in strongly disordered systems, of interest to us here, the bias disorder is much larger than the tunneling amplitudes\cite{SS09}. Calculation of the bias energy can therefore be carried out without consideration of the tunneling. The tunneling amplitude, which for almost all TTLSs is much smaller than the bias energy, can then be considered after the distribution of biases, which is of interest for us here, has been established.
We thus start by orienting each $CN$\textsuperscript{-} ion in one of the inspace diagonals (plane diagonals in 2D) randomly.
We then relax the system for the given initial orientation of the $CN$\textsuperscript{-} impurities to the local energy minimum, optimizing all ion locations as well as $CN$\textsuperscript{-} orientations using the non-linear Fletcher-Reeves conjugate gradients method. Periodic boundary conditions are used to simulate the infinite crystal.

The conjugate gradient method, however, does not allow the crossings of the high barriers separating different single CN states. Such crossings are required, though, to find a low energy state of the full system, as is dictated by the impurity-impurity interactions. We thus combine the conjugate gradients method with intervening MC steps. After the initial energy minimization, the orientation of one $CN$\textsuperscript{-} ion is changed randomly to another in space diagonal (in plane diagonal in 2D), and the system's energy is minimized again using conjugate gradient method. The new configuration is accepted or rejected according to the standard Metropolis algorithm and the assigned temperature. These steps are carried at 40 different temperatures from $300$K to $0.02$K, with 256 MC steps for each temperature, where a single MC step involves a single orientation change of each $CN$\textsuperscript{-} ion in the sample.

We then do 256 MC steps at $10^{-5}$ K, and reach a low energy metastable state. The final state of the system is not necessarily the ground state. However, such metastable states are expected to represent well the real low temperature glassy state of KBr:CN and its low energy DOS. The latter is dictated by the Efros Shklovskii gap\cite{ES75}, which relies only on stability to single and double $CN$\textsuperscript{-} flips or rotations\cite{BEGS79}.

Once we reach the final state of the simulation we measure all single particle excitation energies. We also keep track of the symmetry of the excitations. The symmetric excitations, i.e. those where the final orientation of the $CN$\textsuperscript{-} ions is nearly inverse to its initial orientation (within $14^\circ$) are denoted
$\tau$-TTLSs, and the asymmetric excitations are denoted $S$-TTLSs\cite{SS09}.
This procedure is then repeated for 3000 samples in 3D and 5000 samples in 2D.

\begin{figure}
\subfigure[]{
  \includegraphics[width=8.5cm]{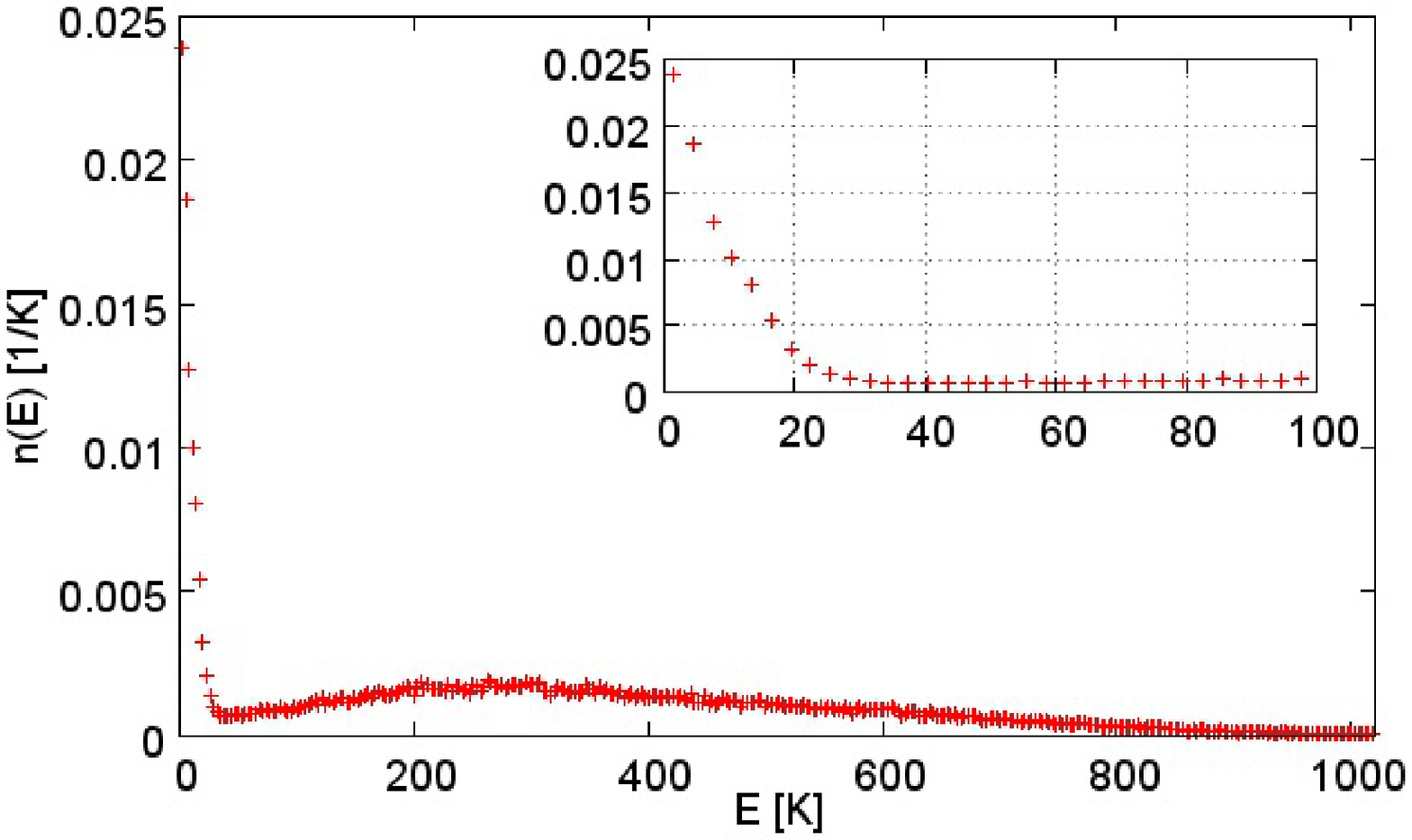}
   \label{fig:Fig1subfig1}
   }
 \subfigure[]{
  \includegraphics[width=8.5cm]{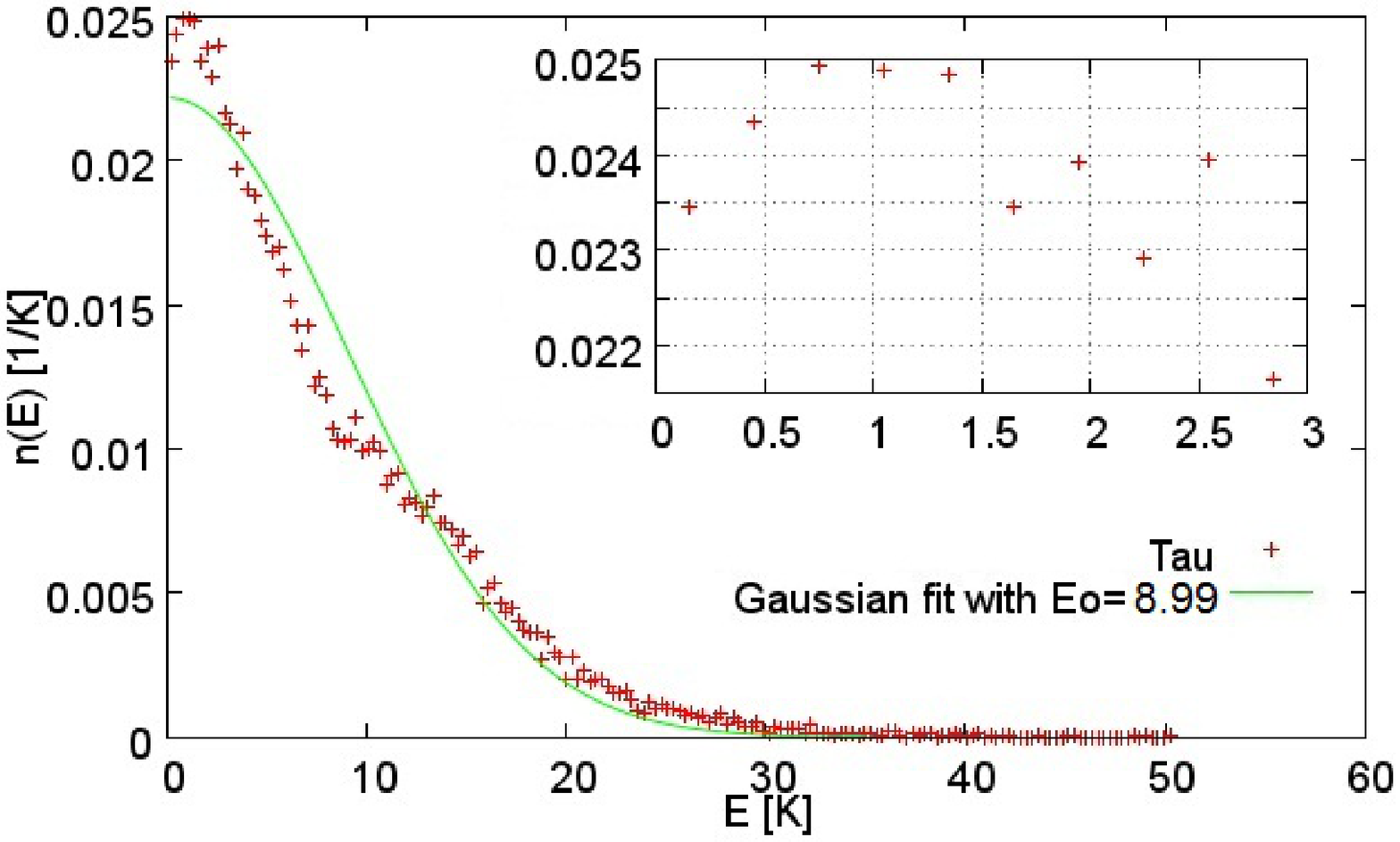}
   \label{fig:Fig1subfig2}
   }
 \subfigure[]{
  \includegraphics[width=8.5cm]{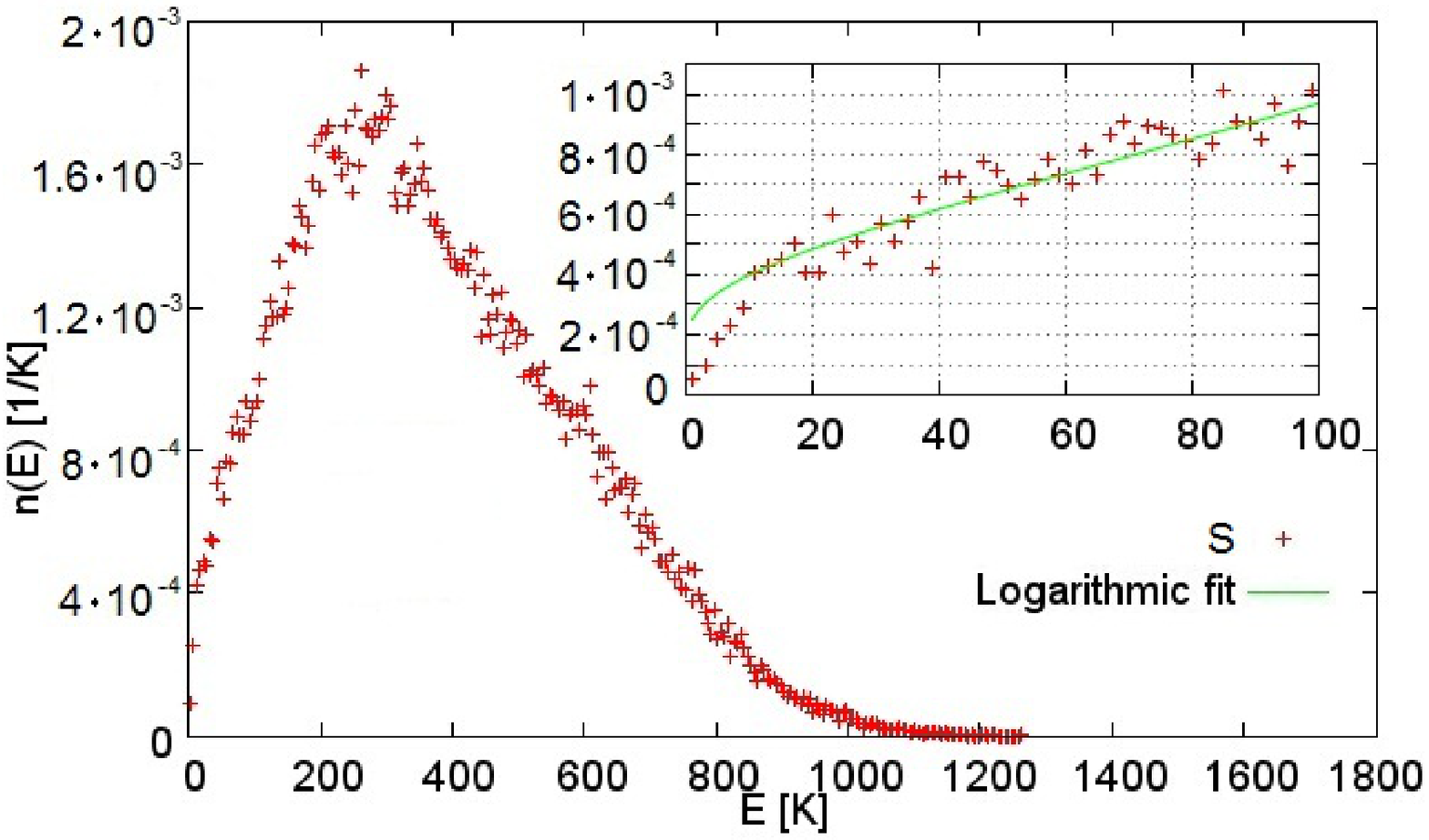}
   \label{fig:Fig1subfig3}
   }
\caption{ The DOS of single impurity tunneling in 3D in 4x4x4 unit cell with CN concentration x=0.25. (a)  Full DOS up to $1000$K. (b) DOS of CN flips ($\tau$-excitations). Solid line denotes a one parameter Gaussian fit, with standard deviation $E_o=8.99$. (c) DOS of CN rotations ($S$-excitations). Solid line in the inset denotes a fit to the function $A/ln(B/E)$ ($A=1.56 \times 10^{-3}, B=500$).   }
\label{fig:DOS3D}
\end{figure}

\begin{figure}
\subfigure[]{
  \includegraphics[width=8.5cm]{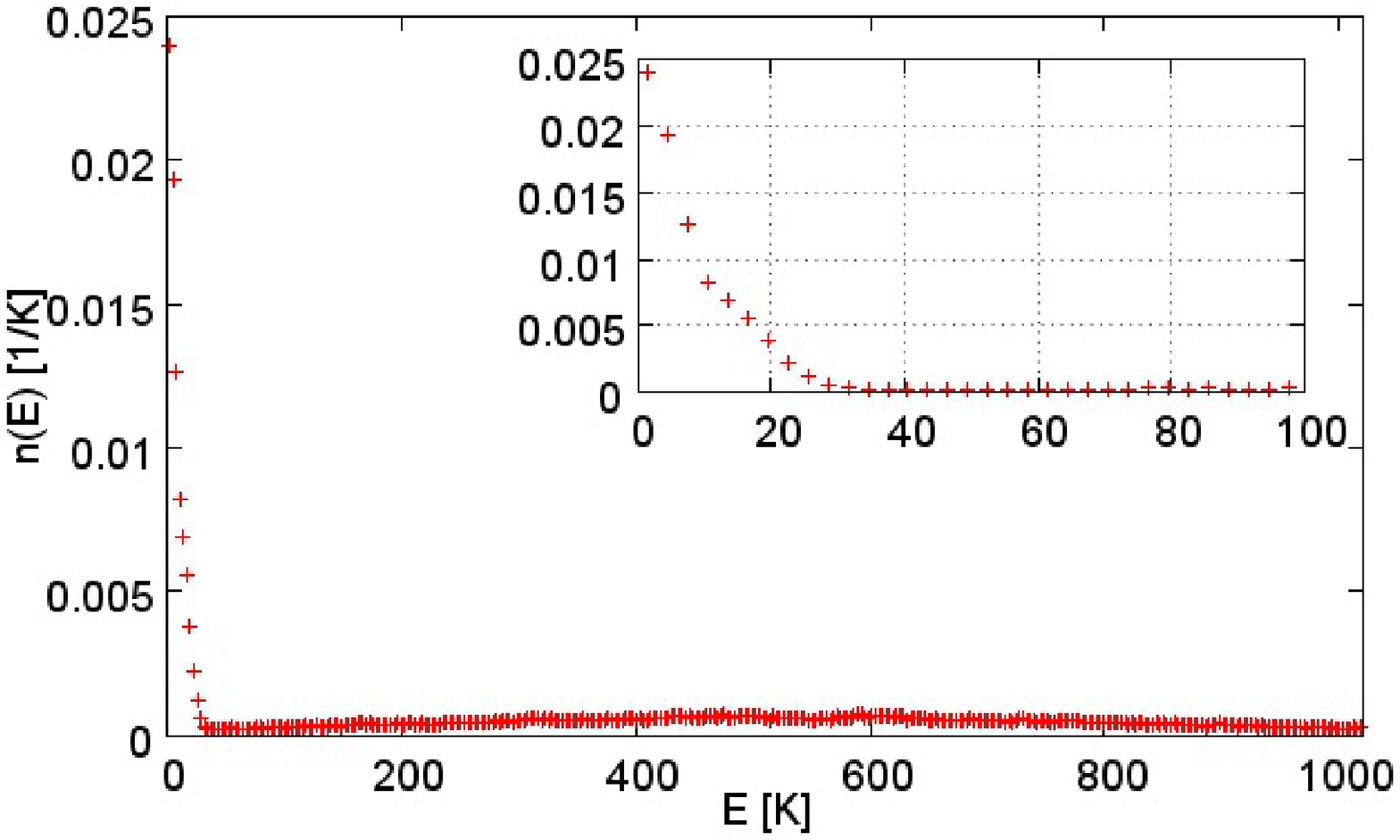}
   \label{fig:Fig2subfig1}
   }
 \subfigure[]{
  \includegraphics[width=8.5cm]{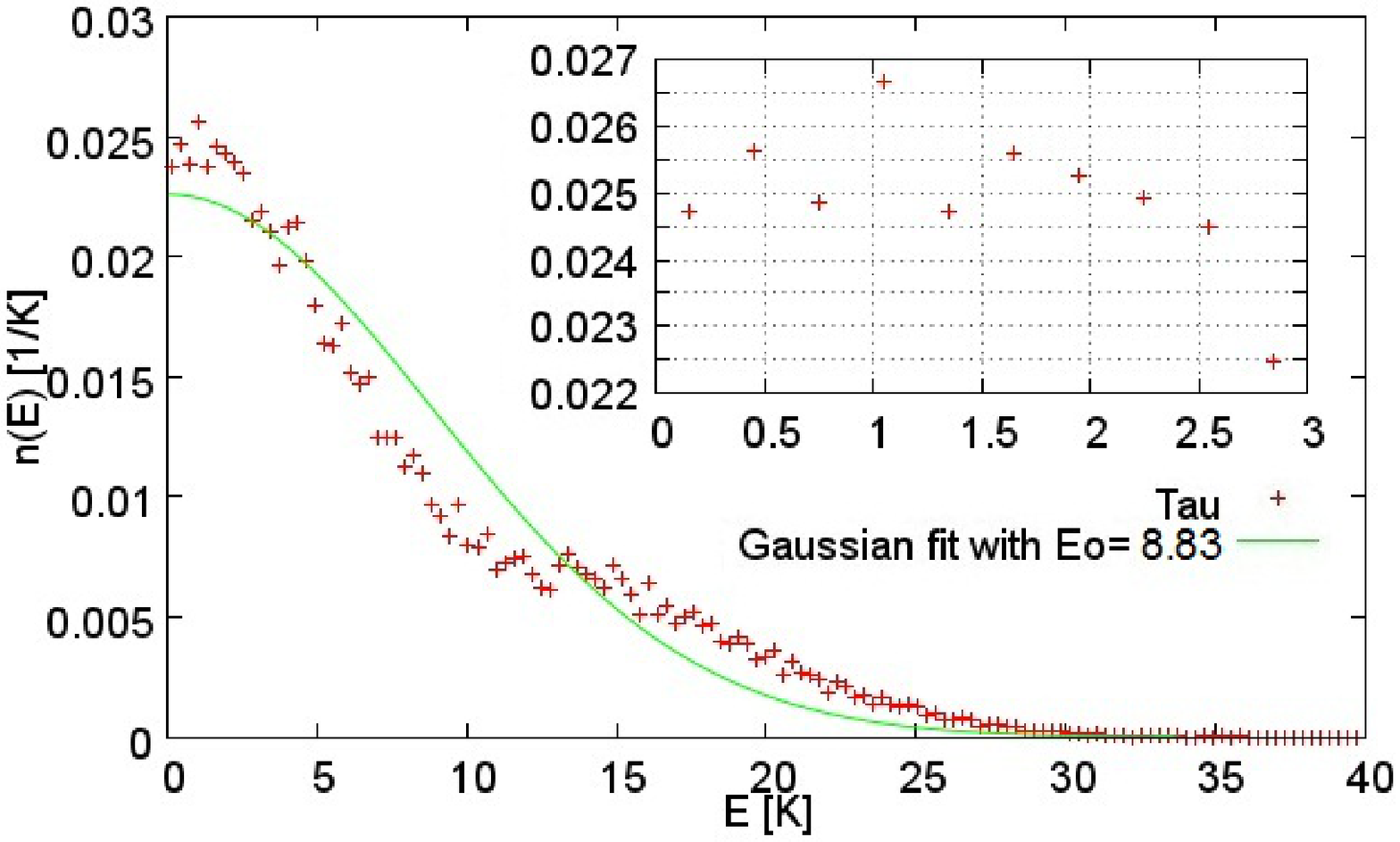}
   \label{fig:Fig2subfig2}
   }
 \subfigure[]{
  \includegraphics[width=8.5cm]{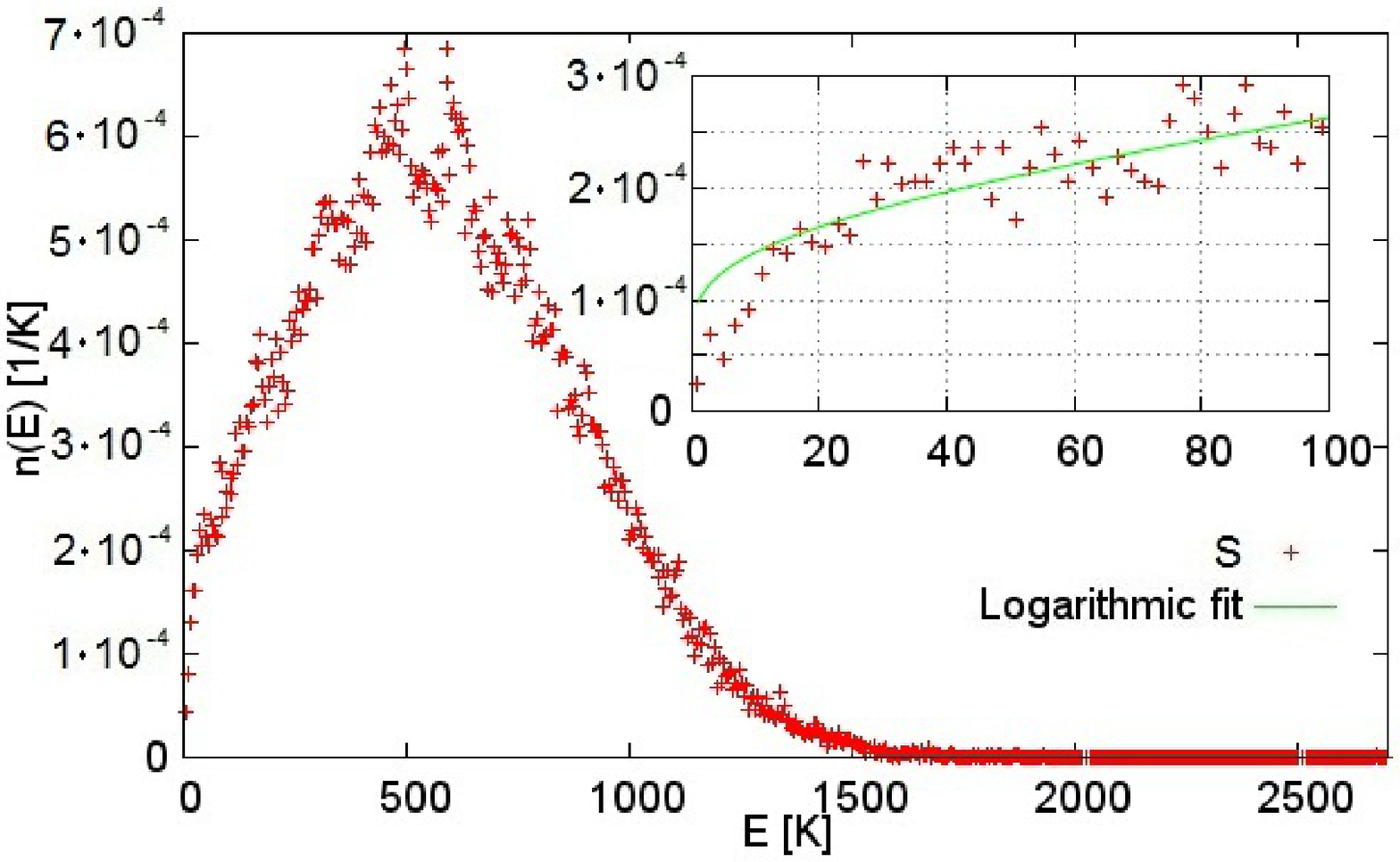}
   \label{fig:Fig2subfig3}
   }
\caption{ The DOS of single impurity tunneling in 2D in 8x8 unit cell with CN concentration x=0.25.  (a)  Full DOS up to $1000$K. (b) DOS of CN flips ($\tau$-excitations). Solid line denotes a one parameter Gaussian fit, with standard deviation $E_o=8.65$. (c) DOS of CN rotations ($S$-excitations). Solid line in the inset denotes a fit to the function $A/ln(B/E)$ ($A=7.7 \times 10^{-4}, B=1500$).   }
\label{fig:DOS2D}
\end{figure}

\section{Results}
\label{Sec:Results}

In Fig.\ref{fig:DOS3D}(a) and Fig.\ref{fig:DOS2D}(a) we plot the full DOS of single impurity tunneling in 3D ($4\times4\times4$ unit cells) and 2D ($8\times8$ unit cells) respectively. The DOS has a maximum at low energies, but reduces considerably at an energy $\approx 3$K.

We now plot separately the single particle excitations of CN flips ($\tau$-TTLSs) and rotations ($S$-TTLSs)
in Fig.\ref{fig:DOS3D}(b),(c) for the 3D samples and in Fig.\ref{fig:DOS2D}(b),(c) for the 2D samples. The difference between the DOS of the two types of excitations is striking. The inversion symmetric flip excitations are peaked at low energies, with typical energies of $\approx 7$K for both 3D and 2D. The inversion non symmetric rotation excitations, however, are broadly distributed, with typical energies of $\approx 300$K for 3D and $\approx 500$K for 2D. The peak value of the $\tau$-TTLS DOS is $\sim 14$ and $\sim 40$ times larger than the peak value of the $S$-TTLS DOS in 3D and 2D respectively. The difference between the peak values of the $S$-TTLSs in 3D and in 2D is a consequence of each CN impurity having two $S$ excitations in 2D, but six $S$ excitations in 3D.

Focusing on the CN flips ($\tau$-TTLSs), their DOS is quite close to a Gaussian, except for a small dip in the DOS at very low energies [see inset of Figs.~\ref{fig:DOS3D}(b),~\ref{fig:DOS2D}(b)], and some structure at $\sim 10$ K. The former is generic to the two-TTLS model (see below) whereas the latter is a result of details of the lattice structure and form of the interaction. With regard to the $S$-TTLSs, their DOS is well fit to a log function at low temperatures, except below $10$ K, the energy at which the $\tau$-TTLSs appear, see further discussion below.

\subsection{Two-TTLS model}
\label{Sec:TwoTLS}

Our results for the DOS are intimately connected with the fact that the coupling strength to the phonon field of inversion symmetric excitations $\gamma_{\rm w}$ is much smaller than that of the asymmetric excitations $\gamma_{\rm s}$. For KBr:CN $\gamma_{\rm w} \approx 0.1eV$, whereas $\gamma_{\rm s} \approx 3eV$\cite{GS11}. This
bi-modality in the coupling strengths affects the phonon mediated acoustic interactions, which then result in an effective TTLS-TTLS interaction Hamiltonian of the form\cite{CBS13,SS09}

\begin{equation}
H_{\rm S\tau}^{\rm eff} = \sum_{ij} U_{ij}^{SS} S_i^z
S_j^z + \sum_{ij} U_{ij}^{S \tau} S_i^z
\tau_j^z + \sum_{ij} U_{ij}^{\tau \tau} \tau_i^z
\tau_j^z
 \label{effectiveStau}
\end{equation}
where in three dimensions (two dimensions) all interactions decay as $1/r^3$ ($1/r^2$) at distances $r \gg a_0$, with a short distance cutoff $\tilde{a} \sim a_0$, and their typical values at near neighbor distance are related by\cite{SS08b,SS09,CBS13}

\begin{equation}
U_0^{\tau \tau} \approx g U_0^{S \tau} \approx g^2 U_0^{S S} .
\label{gratios}
\end{equation}
Here $g \equiv \gamma_{\rm w}/\gamma_{\rm s} \approx 1/30$\cite{SS09,GS11}. This effective Hamiltonian was shown to lead to quantitative universality and to account for the energy scale of $\approx 0.2 U_0^{S \tau} \approx 0.2gU_0^{SS} \approx 3$K related to it\cite{SS09}. The DOS for the two types of TTLSs within the Hamiltonian~\ref{effectiveStau} was analyzed in detail in Ref.\cite{CGBS13}. Indeed, it was found that the weakly interacting ($\tau$) TTLSs DOS are well fit by a Gaussian, except for a small dip, at very low energies, of relative magnitude $\sim g$. The Gaussian width is dictated by the strength of the $S-\tau$ interaction $U_0^{S \tau}$. The strongly interacting $S$-TTLSs were found to have a typical energy dictated by $U_0^{SS}$, a logarithmic gap at intermediate energies\cite{BSE80,Bur95}, and a power-law gap below $\approx U_0^{S \tau}$, the energy where the DOS of the $\tau$-TTLSs becomes appreciable. As a consequence of this gapping of the $S$-TTLS DOS $n_S(E)$, one finds below $\approx 3$ K that $n_S(E) \gamma_S^2 < n_{\tau}(E) \gamma_w^2$ (i.e. $n_S(E)/n_{\tau}(E) < g^2$). This defines the condition for the $\tau$-TTLS to dominate phonon attenuation, resulting in universality below this energy scale.

Our results here validate the applicability of the above two-TTLS model to real systems, as we verify, in a calculation relying only on the bare interatomic interactions of a real glass, the central features predicted by the model: (i) The width of the distributions (typical energies) of the $S$-TTLSs and $\tau$-TTLSs - both their relative magnitudes and the absolute magnitude of each one. We further note that the value of $g$ inferred from the ratio between the typical energies is in good agreement with that obtained from the ratio of the interaction constants in Ref.\cite{GS11}; (ii) The gapping of the $S$-TTLSs but not the $\tau$-TTLSs at low energies, and the small dip, of relative magnitude $\sim g$, of the $\tau$-TTLSs at the lowest energies: (iii) The change in the functional form of the gap of the $S$-TTLSs at the energy scale $\sim 10$ K where the $\tau$-TTLSs appear: (iv) The relation $n_S(E)/n_{\tau}(E) < g^2$ is fulfilled at energies smaller than $\sim 3$ K. We note that the energy below which the above condition is fulfilled in KBr:CN is slightly larger than what is inferred by our results, because as a consequence of finite size effect our results over estimate the $S$-TTLS DOS at the lowest energies.

\section{Discussion}
\label{Sec:Discussion}

In contrast to the assumption of the STM, we find that in KBr:CN TTLSs are divided to two distinct classes by their local inversion symmetry, with a DOS distribution which is particular to each class of TTLSs, and has a strong energy dependence at low energies, of order $10$ K. Many disordered lattices, of which KBr:CN is a primary example, share the low temperature universal characteristics with amorphous solids. Furthermore, experiments strongly support the notion that it is the same mechanism, pertaining to the disordered state of matter itself, that dictates universality in amorphous systems and disordered lattices alike. The structure we find for the DOS of the symmetric and asymmetric TTLSs is not only very different from the flat DOS suggested by the STM, but it is this very structure of DOS that provides an explanation of the quantitative universality of phonon attenuation and the energy scale of $\approx 3$K below which universality is observed\cite{SS09}. As it is expected that the same model would be relevant to both amorphous solids and disordered lattices showing universality, our results suggest that also in amorphous solids two types of TTLSs exist, with a similar structure of DOS to the one found here. Verification of this idea could lead not only to a resolution of the long standing problem of the low temperature universality in disordered systems, but also to an enhanced understanding of the microscopic structure of amorphous solids, and its relation
to the physical properties of amorphous solids in general.

The standard tunneling model uses only one feature of the complicated structure of the density of states found here - the rather homogeneous DOS of the inversion symmetric TTLSs below $3$K. It completely neglects not only the sharp reduction of the DOS at higher energies, but also the existence of the asymmetric TTLSs with their much larger interaction with the phonon field. These characteristics of the DOS, however, may lead to the explanation of further intriguing properties of disordered lattices and amorphous solids at low and intermediate temperatures. Some examples at intermediate temperatures are the full temperature dependence of the thermal conductivity including the rather universal plateau between $3-10$K\cite{YKMP86,FA86}, and the boson peak. At low temperatures, the existence of the strongly interacting $S$-TTLSs, although small in number, may be useful in explaining equilibrium and non-equilibrium properties of phonon attenuation, as well as the relaxation and decoherence of the prevalent $\tau$-TTLSs.

In addition to the standard TTLSs responsible to the universal phonon attenuation at low temperatures, a second type of local excitations was introduced previously, phenomenologically, by Black and Halperin\cite{BH77} and by Yu and Freeman\cite{YF87}. Black and Halperin suggested the existence of TTLSs with a much smaller interaction with the strain, thus affecting the specific heat but not phonon attenuation. Yu and Freeman suggested the existence of local phononic modes with a similar interaction with the phonon field to that of the TTLSs, but with a DOS which is gapped below $43$ K, to fit experimental data for thermal conductivity at intermediate and high temperatures. Our results differ from the above two approaches in that TTLSs of the second type have a much larger interaction with the strain, and are softly gapped below $\sim 3$ K.

Experimentally, TTLSs are observed via a plethora of techniques, including echo\cite{GG76,GSHD79,NFHE04}, coupling to superconducting qubits\cite{LSC+10,GPL+12}, specific heat\cite{DFA86} and dielectric response\cite{EWNS88}. In general, our results suggest that care should be taken in experiments to characterize the TTLSs according to their interaction strength. Whereas most experiments are susceptible to the weakly interacting TTLSs abundant at low energies, strongly interacting TTLSs can dominate experimental observations susceptible to their high bias energy or strong interaction with the strain. Specifically, our results disagree with previous estimates for typical energy biases of CN flips in KBr:CN\cite{SC85,SK94}, and suggest that experimental values of $300-400$ K obtained for TTLS-TTLS interaction at short distances, and for TTLS bias energies, in KBr:CN\cite{DFA86,EWNS88} concern CN rotations rather than CN flips.

The use of the empirical potential of the (Buckingham-Coulomb) form given in Eq.~(\ref{potential}) rather than the more accurate DFT or ab-initio formalisms is dictated by the required complexity of the problem: the need to consider a relatively large system, to find its low energy state using Monte Carlo, and to repeat the calculation for many samples for good statistics. This use of the empirical potential has been shown to be useful in obtaining qualitative, and to some degree also quantitative, predictions for ionic crystals (for details see Refs.~\cite{BK82,KM83,LC85,HUH+05}).
For the properties which are of interest to us here, i.e. the DOS of the $CN$\textsuperscript{-} flips ($\tau$-TLSs) and $CN$\textsuperscript{-} rotations ($S$-TLSs), all qualitative features are dictated by symmetry and are therefore robust. Comparison to results for the model Hamiltonian in Eq.(\ref{effectiveStau})\cite{GS11,CGBS13} suggest that all qualitative features of the DOS for both $CN$\textsuperscript{-} flips and $CN$\textsuperscript{-} rotations  are indeed retrieved by our calculation.
Quantitatively, the value of $g$ inferred from our results for the typical energies of $CN$\textsuperscript{-} flips and $CN$\textsuperscript{-} rotations is in very good agreement with that found by the ratio of their corresponding couplings to the phonon field using DFT and ab-initio calculations\cite{GS11}. We also find good quantitative correspondence with the calculations for the model Hamiltonian in Eq.(\ref{effectiveStau}) as obtained in Ref.~\cite{CGBS13}. This suggests that our results provide also reasonable quantitative accuracy, except for the over estimate of the DOS of the $S$-TLSs at the very low energies, as mentioned above.

We are concerned here with the calculation of the bias energies of the $S$-TLSs and the $\tau$-TLSs, but not their tunneling amplitudes. The latter are crucial for the discussion of dynamic and non-equilibrium properties. A plausible assumption for the distribution of the tunneling amplitudes $\Delta_0$ is that suggested in Refs.\cite{AHV72,Phi72}, $P(\Delta_0) \propto 1/\Delta_0$. For the $\tau$-TLSs which dominate the low energy universal phenomena, this assumption has been shown to be consistent with the various experimental findings. However, its applicability for the $S$-TLSs needs to be checked. Such a verification may become possible by using our results for the bias energies here, and for the coupling constant of the $S$-TLSs in Ref.\cite{GS11} to calculate measurable quantities dominated by $S$-TLSs and comparing to experimental results.


{\it Acknowledgments} ---
We thank A. Burin, A. Gaita-Ari\~no, Niels Gr{\o}nbech-Jensen, and Joerg Rottler for very useful discussions. This research was supported by the Israel Science Foundation (Grant No. 982/10).

\end{document}